\documentclass[12pt]{iopart}
\usepackage{iopams}
\usepackage{setstack}
\usepackage{epsfig}
\usepackage{rotating}
\begin{document}

\title{A matched expansion approach to practical self-force calculations}
\author{Warren G. Anderson and Alan G. Wiseman}
\address{Department of Physics, University of Wisconsin --
Milwaukee, P.O. Box 413, Milwaukee, Wisconsin, 53201, U.S.A.}
\ead{\mailto{warren@gravity.phys.uwm.edu},\mailto{agw@gravity.phys.uwm.edu}}

\begin{abstract}%
We discuss a practical method to compute the self-force on a particle moving
through a curved spacetime. This method involves two expansions to calculate
the self-force, one arising from the particle's immediate past and the other
from the more distant past. The expansion in the immediate past is a covariant
Taylor series and can be carried out for all geometries. The more distant
expansion is a mode sum, and may be carried out in those cases where the wave
equation for the field mediating the self-force admits a mode expansion of the
solution.  In particular, this method can be used to calculate the
gravitational self-force for a particle of mass ${\mu}$ orbiting a black hole of
mass $M$ to order ${\mu}^2$, provided ${\mu}/M {\ll}1$. We discuss how to use these two
expansions to construct a full self-force, and in particular investigate
criteria for matching the two expansions. As with all methods of computing
self-forces for particles moving in black hole spacetimes, one encounters 
considerable technical difficulty in applying this method; nevertheless, it
appears that the convergence of each series is good enough that a practical
implementation  may be plausible. 
\end{abstract}
\submitto{\CQG}
\maketitle

\section{Introduction} 
\label{s:intro}

There is currently considerable interest in calculating the self-force on a
charged particle moving in a curved background spacetime. This interest is
twofold. On one hand, self-force is an intrinsically interesting phenomenon,
apparently both deep and subtle. On the other, there is increasing practical
interest in understanding the motion of particles with small but
non-negligible masses in curved geometries. In particular, there is
wide-spread belief that accurate computation of gravitational self-force is
needed to calculate gravitational waveform templates for key data analysis
efforts associated with interferometric gravitational wave detectors. With
such templates, the Laser Interferometer Space Antenna (LISA)\cite{LISA} is
expected to produce a wealth of information about strongly-field
gravity\cite{Ryan96}.

We define self-force to be any force on a particle of quadratic (or higher)
order in the charge carried by that particle. For electrically charged
particles in flat spacetime, the self-force is given by the
Abraham-Lorentz-Dirac formula \cite{Dirac38,Poisson99}.  DeWitt and
Brehme\cite{DB60} derived the general expression for the self-force on an
electrically charged particle in a curved background (however, they missed a
term which was later supplied by Hobbs\cite{Hobbs68}).  The gravitational
self-force was first calculated almost simultaneously by Mino, Sasaki and
Tanaka \cite{MST97} and by Quinn and Wald \cite{QW97}.  Later, Quinn derived
the equivalent formula for a charge coupled to a minimally-coupled massless
scalar field\cite{Quinn00}.  These results have resolved many of the issues of
principle in computing self-forces in curved spacetime.
Poisson\cite{Poisson03} has recently written a comprehensive review article in
which these expressions and their implications are discussed at length.

The self-force expressions for all fields of physical interest are of similar
form. We henceforth restrict our attention, therefore, to the simplest case;
a particle with mass ${\mu}$ and scalar charge $q$ coupled to a minimally-coupled
massless scalar field in a curved background geometry. Lessons learned here
should be extendible to other physical fields without major modification.

The field equation for a minimally-coupled massless scalar field ${\phi}$ is
\begin{equation}
   {\square}{\phi}=-4{\pi}{\rho}. \label{eq:SFE}
\end{equation}
Here, ${\square}$ is the D'Alembertian of the curved background and ${\rho}$ is the charge
density.  We consider a point particle, in which case
\begin{equation}
   {\rho}(x)={\int}q~{\delta}^4(x,z({\tau}))d{\tau}, \label{eq:ChargeDens}
\end{equation}
where ${\delta}^4({\cdot})$ is a generalized Dirac distribution in four dimensions and
$z({\tau})$ denotes the worldline of the particle. 

For such a particle, Quinn\cite{Quinn00} has shown that the self-force is
given by
\begin{eqnarray}
   \fl f^{\alpha}=q^2\left[\frac{1}{3}\left(\dot{a}^{\alpha}-a^2\,u^{\alpha}\right)+
      \frac{1}{12}\left(2\,R^{{\alpha}{\beta}}u_{\beta}+2\,R_{{\beta}{\gamma}}u^{\beta}u^{\gamma}u^{\alpha}-R\,u^{\alpha}\right) 
      \right.\nonumber\\ 
      \lo + \left.\lim_{{\epsilon}{\rightarrow}0^+} {\int}_{-{\infty}}^{{\tau}-{\epsilon}}{\nabla}^{\alpha} G_{\rm ret}
      \left(z({\tau}),z({\tau}')\right) d{\tau}'\right].
\label{eq:QSF}
\end{eqnarray}
The quantities in equation (\ref{eq:QSF}) are defined as follows: $u^{\alpha}$ is the
four-velocity of the particle and  $a^{\alpha} = u^{\beta}{\nabla}_{\beta} u^{\alpha}$ its four-acceleration.
$\dot{a}^{\alpha} = u^{\beta}{\nabla}_{\beta}a^{\alpha} $ denotes derivative of the acceleration with respect
to its proper time, and $a^2=a^{\alpha} a_{\alpha}$ is the magnitude of the acceleration
squared.  The quantity
$R_{{\alpha}{\beta}}$ is the Ricci tensor of the background
spacetime, and $R$ is its scalar curvature.  ${\tau}$ is the
proper time of the particle at its current position, while
${\tau}'$ denotes the proper time at any other point along the
particles worldline.  Finally, $G_{\rm ret}(x,x')$ is the
retarded Green's function for the scalar field equation (\ref{eq:SFE}), which
satisfies the Green's function equation
\begin{equation}
   {\square}~G_{\rm ret}~(x,x') = -4{\pi}{\delta}^4(x,x'), \label{eq:GRetDef}
\end{equation}
and has support only when $x'$ is in the causal past of $x$.

Notice that the terms on the right-hand-side of (\ref{eq:QSF}) can be gathered
into three groups. The first group are local terms involving the acceleration
of the particle. These give the Abraham-Lorentz-Dirac force on an accelerating
particle in flat spacetime\cite{Dirac38}. They arise because the accelerating
particle radiates, and that radiation produces a ``radiation reaction'' or
recoil force on the particle. It is interesting to note that for a particle in
Minkowski spacetime with constant $a$ (whose worldline is a hyperbola), that
$\dot a^{\alpha}\ne0$ (because the direction of the four-vector changes). In fact,
$a^{\alpha}= a^2 u^{\alpha}$ in this case, and the first group vanishes. This fact, which
might seem surprising at first, can be understood as a consequence of the
equivalence principle\cite{Rorlich63,Boulware80,Matsas94}. We further note that for a particle freely falling in
curved space (i.e. following a geodesic), $a^{\alpha}=0$, and thus these terms will
also vanish. 

The second group are also local terms,
this time involving the background curvature. These terms involve
only the Ricci curvature, and thus represent a self-force
mediated by the matter content of the background.  Interestingly,
it is not necessary for the matter to interact directly with the
particle for this to be true. We will largely ignore these first
two groups of terms for the remainder of our discussion because
for a particle following a geodesic in a vacuum background
spacetime, which is the case of most practical interest, they
vanish. Moreover, even when they do not vanish, they are easily
calculated.

In contrast, there is considerable practical difficulty in calculating the
single non-local term which constitutes the third group. Interestingly, this 
non-local self-interaction arises in curved backgrounds, but not in a flat
background. This is because of two ways in which the propagation of massless 
fields differs in curved and flat backgrounds. In Minkowski space, massless
fields propagate along null geodesics. Thus, a particle would have to be
null-separated from some point on its past worldline to affect itself.
However, the simple causal structure of Minkowski space does not allow this
for a massive particle. The particle is restricted to a time-like geodesic,
and no two points in Minkowski space can be connected both by a time-like
geodesic and a null geodesic. Thus, any point with which the charged particle
can interact, it cannot travel to, and vice versa. This is not true in a
curved background, however, where the causal structure can be considerably
more complicated. In this case, it is possible for the field, which ``leaves''
the particle along a null geodesic, to re-intersect that particle, which
follows a timelike geodesic, at a later time.

If this were the only mechanism by which non-local self-interactions arose,
then there would be no self-interactions arising from within the normal
neighbourhood of the point at which the particle sits. Recall that the normal
neighbourhood of a point $x$ is the set of all other points which are
connected to $x$ via a unique geodesic. Recall also that such a neighbourhood
is guaranteed to exist. Thus, if a particle at $x$ is massive and
non-accelerating (i.e. following a timelike geodesic), then any point on the
particle's worldline connected to $x$ by a null geodesic is connected to $x$
by (at least) two geodesics. Therefore, by definition, it is not in the normal
neighbourhood of $x$. 

However, in general, fields do not propagate only along null geodesics in
curved backgrounds. This is apparent from the Hadamard\cite{Hadamard} form of
the retarded Green's function,
\begin{equation}
 \fl  G_{\rm ret}(x,x')={\Theta}[t-t']
      \Bigl\{U(x,x')~ {\delta}[{\sigma}(x,x')] -V(x,x')~ {\Theta}[-{\sigma}(x,x')]\Bigr\},
\label{eq:Hadamard}
\end{equation}
where ${\Theta}[{\cdot}]$ is the Heaviside step function, ${\delta}[{\cdot}]$ is the Dirac delta
distribution, ${\sigma}(x,x')$ is one half of the square of the geodesic distance
between points $x$ and $x'$, and $U(x,x')$ and $V(x,x')$ are smooth functions
which depend on the details of the background and field equation. Note that
this expression is only well defined if one can unambiguously define geodesic
distance between $x$ and $x'$, which implies that $x'$ is within the normal
neighbourhood of $x$. 

Recall that the Green's function gives the field at position $x$ due to a
charge at position $x'$. Now, the first term on the right-hand-side of
equation (\ref{eq:Hadamard}), which is known as the \textit{direct part}
of the Green's function, has support only when the geodesic distance between
$x$ and $x'$ vanishes, or, in other words, when $x$ and $x'$ are separated by
a null geodesic. This term is always present, but, as described above, cannot
contribute to the self-force. However, the second term, which is known as
the \textit{tail} part of the Green's function, does contribute whenever the
square of the geodesic distance between $x$ and $x'$ is negative, or, in other
words, when $x$ and $x'$ are separated by a timelike geodesic\footnote{This
fascinating fact was first elaborated upon by Hadamard\cite{Hadamard}, who
described it as a failure of Huygens' principle to hold in general for
hyperbolic partial differential equations - c.f.  the Paul G\"{u}nther
memorial edition of Zeitschrift f\"{u}r Analysen und ihre
Anwendungen\cite{Gunther} for a collection of articles addressing the proof of
the modified Hadamard's conjecture, which states that the only spacetimes for
which standard wave equations obey Huygen's principle are those conformal to
Minkowski spacetime and one plane-wave family of spacetimes.}. Since every
point on the particles past worldline is timelike separated from the particle
by a timelike geodesic (to wit, the worldline), through this term a self-force
can be generated at every time within the normal neighbourhood. 

Given that all points on a particles past worldline can interact with the
particle, it may seem somewhat arbitrary to have singled out in the above
discussion points on the past worldline which are null-separated from the
particle.  Indeed, in the literature, it is often stated that the self-force
arises from the tail part of the field (or Green's function) and left at that.
This statement can be somewhat confusing, however. Outside of the normal
neighbourhood, there is no clear distinction between tail and direct parts - is
the field from a point connected by both timelike and null geodesics a direct
field, a tail field, or both?  

\begin{figure} 
   \begin{center} 
      \epsfysize=7.5cm
      \epsfbox{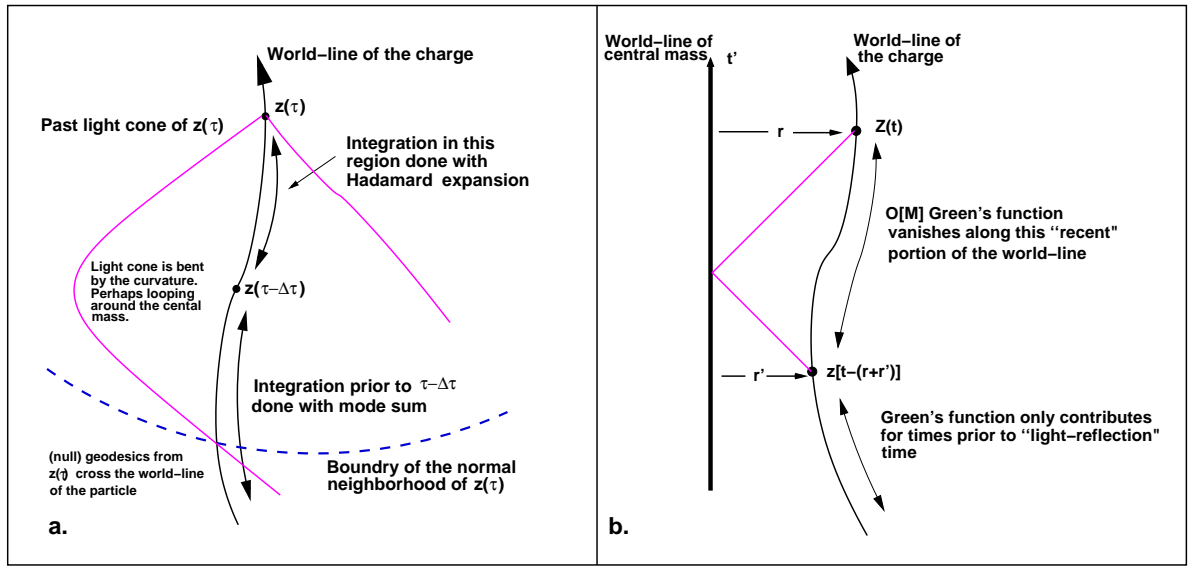} 
   \end{center}
   \caption{
   In order to compute the force at $z({\tau})$, one needs to
integrate back over the entire history of the particle.  Figure
\ref{fig:merged}.a depicts the basic idea behind this paper: use
the Hadamard expansion to integrate over the portion of the
worldline ``near"  $z({\tau})$, and then finish off the integral
using a mode sum for points well separated from (and extending
beyond the normal neighbourhood  of) $z({\tau})$.  The choice of when to
do the switch is also alluded to in the figure: the boundary of
the normal neighbourhood -- the crossing of geodesics -- will
likely be due to null geodesics looping around the black hole and
re-intersecting the particle's worldline.  This is discussed in
Section \ref{sec:Anderson} and the results are given in Table
\ref{tab:pt}.  It is probably good to do the switch somewhere in
the middle, where keeping only a few terms in the Hadamard
expansion is still a good approximation and before the mode sum
is fighting against the divergent quantities on the light
cone.  Figure \ref{fig:merged}.b shows that for the $O[M]$
Green's function (see section \ref{sec:Wiseman}) it is actually
the portion of the past trajectory that is well removed from the
current position (i.e. $t^{\prime}\le t - (r+r')$) that actually 
produces the ``tail" force.
Rigorously, it make little sense to talk of a ``normal
neighbourhood" for a spacetime with weak curvature, yet the
similarity in figure \ref{fig:merged}.a and figure \ref{fig:merged}.b 
strongly suggests that the portion of the worldline near 
the normal neighbourhood boundary and outside the normal
neighbourhood boundary dominate the contribution  to the self
force.
   } 
   \label{fig:merged}
\end{figure}

Furthermore, there is reason to believe that these points play a
qualitatively different role in the self-interaction. In
particular, when doing the integral over the Green's function,
distributions (i.e.  ${\delta}$-functions and step functions) will
be encountered when the source point and field point are null
separated. In effect, the particle can ``feel" its own direct
field ``sent"  from points in the past. This is depicted in
figure \ref{fig:merged}. In the specific case of the $O[M]$
Green's function, such distributions do appear, and their
contribution make up the entire self-force  (see equation
(\ref{eq:OMGFanalytic})).


Let us turn now to the matter of calculating the non-local part of the
self-force,
\begin{equation}
     f^{\alpha}=q^2 \lim_{{\epsilon}{\rightarrow}0^+} {\int}_{-{\infty}}^{{\tau}-{\epsilon}}{\nabla}^{\alpha} G_{\rm
    ret}\left(z({\tau}),z({\tau}') \right)d{\tau}' .
   \label{eq:NLSF}
\end{equation}
The key feature of this formula is that the integral is well behaved over the
entire region of integration, i.e. it is integrable throughout its domain.
The termination of the integral at ${\tau}-{\epsilon}$ (as opposed to integrating all the
way to ${\tau}$) is a required for this to be true, however, because the Green's
function has an  unintegrable  singularity at the coincidence limit (${\epsilon}=0$).
Nonetheless, this contribution is distributional, and the integrand is
perfectly well behaved as long as we approach ${\tau}'={\tau}$ from the past. 

At first glance, evaluating equation (\ref{eq:NLSF}) seems like a relatively
straightforward task - an approximate retarded Green's function can be
calculated, for instance, using mode-sum techniques\cite{B&D}.  However, in
practice, we can only sum a finite number of modes and thereby obtain an
approximate Green's function.  Furthermore, as we will demonstrate in section
\ref{sec:Wiseman}, the number of modes needed to obtain a given accuracy grows
without bound as epsilon approaches zero. This does not preclude the use of
modes to calculate the non-local part of the self-force, and, as can be seen
in the pages of this special issue, modes are indeed widely used. The modes
must, however, first be regularized by some method.

Nonetheless, it is our purpose in this paper to explore an alternative method
which may have some advantages over a regularized mode sum. It does not need
to make use of regularized modes, although one might choose to do so.  Rather,
it is a method of matched expansions, in which one calculates the self-force
using  the tail within some portion of the normal neighbourhood of the particle
and using an unregularized mode expansion for the remainder of the
particle's worldline. More precisely, we propose to express equation
(\ref{eq:NLSF}) as 
\begin{equation} \fl f^{\alpha}=-q^2\,
   {\int}_{{\tau}-{\Delta}{\tau}}^{{\tau}}{\nabla}^{\alpha}\,V\left(z({\tau}),z({\tau}')\right)d{\tau}' +q^2\,{\int}_{-{\infty}}^{{\tau}-{\Delta}{\tau}}{\nabla}^{\alpha}\,G_{\rm
   ret}\left(z({\tau}),z({\tau}')\right) \, d {\tau}, 
   \label{eq:PWSF} 
\end{equation} 
where ${\Delta}{\tau}$ is an interval of proper time.  We require that ${\Delta}{\tau}$  be chosen so
that $z({\tau}')$ is within the normal neighbourhood of $z({\tau})$ for all ${\tau}-{\Delta}{\tau}<{\tau}'<{\tau}$.
In other words, ${\Delta}{\tau}$ distinguishes the contribution to the self-force coming
from the \textit{recent history} of the particle from the contribution that
comes from the more distant past.  It is the choice of ${\Delta}{\tau}$ and the
feasibility of evaluating the two integrals that will occupy us for the
remainder of this paper.

There are several notable features of equation (\ref{eq:PWSF}).
First, this expression is exactly equivalent to equation
(\ref{eq:NLSF}), since, as is evident from equation
(\ref{eq:Hadamard}), $G_{\rm ret}(x,x'){\equiv} -V(x,x')$ provided
we restrict ourselves to the interior of the past light cone, as
is always the case for this integral.  Second, there is no longer
a limit needed because $V(x,x')$ is regular everywhere. Finally,
notice that we now have a new parameter, ${\Delta}{\tau}$, which
we are free to choose so long as it is not too large. 

One might suspect that the contribution to the Green's function ``falls off"
fast enough, in general, so that perhaps only the first integral needs to be
evaluated; and, since it is restricted to the normal neighbourhood, only the
Hadamard expansion is needed to compute the self-force.  Surprisingly, there
are simple situations in which this line of reasoning turns out to be false;
the second integral gives a significant (perhaps, the dominant)
contribution to the total force. We demonstrate this in section
\ref{sec:Wiseman} using a Green's function for spacetime with a large central
mass $M$ where we only keep the terms of leading order in the central mass
(the ``O[M] Green's function").  Using the O[M] Green's function, none of the
force originates from the tail contribution in the normal neighbourhood; all
of the force comes times prior to the light reflection time.

Unfortunately, neither integrand in equation (\ref{eq:PWSF}) can be calculated
exactly. As mentioned above, $G_{\rm ret}(x,x')$ can be approximated in the
second integrand by a mode sum expansion. The advantage here is that the mode
sum does not need to be extended to the limit of the particle's position.
Thus, for any fixed finite precision required, a finite number of modes will
be needed for the second integral in equation (\ref{eq:PWSF}). As noted above,
however, that number will grow as ${\Delta}{\tau}$ decreases.

On the other hand, for $V(x,x')$, we can take advantage of the fact that we
are within the normal neighbourhood, and can therefore define a Riemann normal
coordinate system. In such a coordinate system, it is relatively
straightforward to expand $V(x,x')$ in a covariant Taylor series. Such a
series will have coefficients constructed of geometric quantities (notably,
the particle's four-velocity and the curvature of the background) evaluated at
the particles position, and will be an expansion in geodesic distance from
that position. Again, only a finite number of terms will be available, so only
an approximation of $V(x,x')$ can be calculated. In contrast to the mode sum
expansion of $G_{\rm ret}(x,x')$, the covariant Taylor series will, in
general, require more terms to achieve a given accuracy as ${\Delta}{\tau}$
\textit{increases}.

The goal, then, is to calculate both series to sufficient accuracy within
their applicable domains, and to then integrate and sum. We may take advantage
or our parameter, ${\Delta}{\tau}$, to adjust the number of terms required in each series
to achieve the required accuracy. However, both series are non-trivial to
calculate, and the difficulty of calculating each successive term is greater
than for the last. There is, therefore, no guarantee that there is any value
of ${\Delta}{\tau}$ such that the number of terms needed in both series can be calculated
\textit{in practice}.

It is the main goal of the remainder of this paper to investigate if it is
plausible that there would exist a choice of ${\Delta}{\tau}$ which would make this
scheme, first proposed by Poisson and Wiseman\cite{PW97}, viable for practical
calculations. Because of the difficulty of performing both expansions, we will
exploit existing results when possible. Furthermore, we will use the simplest
and most transparent results that still bear upon the problem. We will not
explicitly try to calculate the self-force for any specific geometry or
particle motion, but will, rather, simply investigate the convergence of both
series expansions.

The plan of the remainder of this paper is as follows: we will review and
examine the existing results for the expansion of the $V(x,x')$ in the next
section. Following that, in section \ref{sec:Wiseman}, we will investigate
the mode expansion. Finally, we will discuss our conclusions in section
\ref{sec:Conc}.

\section{The quasi-local part of the self-force}\label{sec:Anderson}
Covariant normal neighbourhood expansions have a venerable history for
calculations of both self-forces\cite{DB60} and quantum field effects in
curved backgrounds\cite{C76,C78}. There is therefore a plethora of material
from which to begin a calculation of the first term in equation
(\ref{eq:PWSF}), which we will call the quasi-local part of the self-force
\begin{equation}
   f^{\alpha}_{\rm QL}= - q^2 {\int}_{{\tau}-{\Delta}{\tau}}^{{\tau}}{\nabla}^{\alpha}\,V\left(z({\tau}),z({\tau}')\right)d{\tau}'.
\end{equation}
There are, however, to our knowledge, only three papers in the literature
which address the calculation of this part of the self-force. The first was by
Roberts, in which he stopped just short of calculating the quasi-local part of
the self-force for an electric charge in an arbitrary
background\cite{Roberts}.  A trivial extension of his paper gives the leading
order to the quasi-local part of the self-force in this case,
\begin{equation}
   f^{\alpha}_{\rm QL} = - \frac{q^2}{4} ({\delta}_{\alpha}^{\beta}+ u_{\alpha}u^{\beta})\,C^{\alpha}{_{{\beta}{\gamma}}}{^{\delta}}{_{;{\delta}}} u^{\beta}
   u^{\gamma}\,{\Delta}{\tau}^2+O({\Delta}{\tau}^3).
\end{equation}
Here, $C_{{\alpha}{\beta}{\gamma}{\delta}}$ is the Weyl curvature tensor of the background at the
particle's position, $u^{\alpha}$ is the particle's four-velocity, and ${\Delta}{\tau}$, as
previously, is a proper time interval along the particle's past worldline.

The second result is due Anderson and Hu\cite{AndHu}. They calculated the tail
of the retarded Green's function in the normal neighbourhood for a particle
with a minimally-coupled massless scalar field in a Schwarzschild background
geometry. By using a Hadamard-WKB expansion for the Euclidean Green's
function, they are able to expand $V(x,x')$ to sixth order in the geodesic
separation. They do not go on to calculate the quasi-local self-force
explicitly (although that is clearly the motivation for their paper), but it
is a relatively straightforward matter to do so. Because their results are
somewhat unwieldy, we refer the reader to their paper for the actual result,
although we will be reproducing the self-force derived from it for two simple
particle motions below.

Most recently, the quasi-local part of the gravitational self-force for a
particle in an arbitrary curved background has been calculated by Anderson,
Flanagan and Ottewill\cite{AFO}. They calculate the first two non-vanishing terms in the
Taylor series, and find 
\begin{eqnarray}
   \fl f_{QL\,{\alpha}}({\tau},{\Delta}{\tau}) = - {\mu}^2 ({\delta}_{\alpha}^{\beta}+ u_{\alpha}u^{\beta}) C_{{\beta}{\gamma}{\delta}{\varepsilon}} C_{{\sigma}\ \, {\rho}}^{\ \, 
      {\gamma}\, \ {\varepsilon}} u^{\delta}u^{\sigma}u^{\rho}{\Delta}{\tau}^2 \nonumber \\ 
   \lo+ {\mu}^2 ({\delta}_{\alpha}^{\beta}+ u_{\alpha}u^{\beta}) u^{\gamma}u^{\delta}\Bigg[\frac{1}{6} C_{{\gamma}{\mu}{\delta}{\nu}} C_{{\varepsilon}\ \, {\sigma}\
      \,;{\beta}}^{\ \, {\mu}\ \, {\nu}} u^{\varepsilon}u^{\sigma} - \frac{3}{20} C_{{\beta}{\gamma}{\mu}{\delta};{\nu}} C_{{\varepsilon}\ \, {\sigma}}^{\ \,
      {\mu}\ \, {\nu}} u^{\varepsilon}u^{\sigma} \nonumber \\ 
   \lo+\frac{1}{3} \left( \frac{1}{2} C_{{\mu}{\nu}{\gamma}{\lambda}} C^{{\mu}{\nu}\ \,{\lambda}}_{\ \ \ {\delta}\ \,\, ;{\beta}}
      + C_{{\mu}{\varepsilon}{\gamma}{\lambda}} C^{{\mu}\ \ \, {\lambda}}_{\ \, {\sigma}{\delta}\ \,\, ;{\beta}} u^{\varepsilon}u^{\sigma}\right) \nonumber \\ 
   \lo-\frac{19}{60}\left( \frac{1}{2} C_{{\mu}{\nu}{\gamma}{\lambda}} C^{{\mu}{\nu}\ \ \, \,;{\lambda}}_{\ \ \ {\delta}{\beta}}
      + C_{{\mu}{\varepsilon}{\gamma}{\lambda}} C^{{\mu}\ \ \ \ ;{\lambda}}_{\ \, {\sigma}{\delta}{\beta}} u^{\varepsilon}u^{\sigma}\right) \Bigg] {\Delta}{\tau}^3 +
      O({\Delta}{\tau}^4).
\label{eq:finalans}
\end{eqnarray}
They also provide explicit expressions for some particle motions in
Schwarzschild and Kerr geometries.

For the purposes of evaluating the convergence of these expansions, we will
focus on the results of Anderson and Hu\cite{AndHu}, both because the scalar
field case is inherently the simplest and because the provide the highest
order expansion with which to work. Unfortunately, Anderson and Hu do not take
the derivative of $V(x,x')$ necessary to calculate the self-force
contribution. Furthermore, their result is expressed in terms of coordinate
expansion rather than an expansion in geodesic distance. It is only for
relatively simple particle motions that one can easily recast coordinate
distance into geodesic distance. Fortunately, it is in the spirit of our
explorations here to take such simple cases, and we shall.

The fundamental question we ask in this section, then, is ``what is the rate
of convergence of the expansion for the quasi-local part the self-force as a
function of ${\Delta}{\tau}$?'' Clearly, as ${\Delta}{\tau}{\rightarrow}0$, the expansion becomes exact at any
order. The more interesting limit is the one where ${\Delta}{\tau}$ approaches the
boundary of the normal neighbourhood. It might be reasonable to expect that the
expansion ceases to converge at all in that case, since that is the boundary
of its domain of validity.  Indeed, there is no fundamental reason that the
domain of convergence of the series expansion could not be much smaller than
the normal neighbourhood. Addressing this question, however, is slightly
complicated by the fact that it is not immediately clear what the value of
${\Delta}{\tau}$ at the boundary of the normal neighbourhood is.

In the case of a Schwarzschild black hole spacetime, at least, one can get
some insight into the normal neighbourhood boundary by studying null geodesics
intersecting a particle in a circular geodesic at radius $R$.  For such a
particle, the angular displacement is related to the particle's proper time by
\begin{equation}
   {\Delta}{\phi}_{\rm particle} = \sqrt{\frac{M}{R-3M}}\,\frac{{\Delta}{\tau}}{R},
   \label{eq:phi_part}
\end{equation}
where $M$ is the mass of the black hole. On the other hand, the angular
deflection (as seen at $r=R$) of a null geodesic passing within a nearest
distance $R_0>3M$ of the black hole is
\begin{equation}
   {\Delta}{\phi}_{\rm photon} =
   2{\int}_{1/R}^{1/R_0}\,du\,\sqrt{\frac{R_0^3}{R_0-2M-R_0^3\,u^2(1-2Mu)}},
   \label{eq:phi_phot}
\end{equation}
(note that if $R_0\le3M$, there is no turning point for the geodesic and it
does not return to larger radii). Finally, the proper time, as measured by the
particle, for a photon to descend from a distance $R$ to within a distance
$R_0$ of the black hole and return to a distance $R$ is given by
\begin{equation}
  \fl {\Delta}{\tau}=2\left(\frac{R_0-2M}{R0}\right) {\int}_{R_0}^R\,dr\,
   \frac{r}{r-2M}\sqrt{\frac{r^3(R_0-2M)}{rR_0(r^2-R_0^2)-2M(r^3-R_0^3)}}.
   \label{eq:pt}
\end{equation}
What we would like to find is the minimum value of $R_0$ (and hence ${\Delta}{\tau}$) for
which a null geodesic intersecting the particle's circular geodesic at a given
$R$ can re-intersect it at a the same $R$ but a different time. For a particle
orbiting at $R{\gg}3M$ from the black hole, the condition for this to occur is 
\begin{equation}
   {\Delta}{\phi}_{\rm particle} + {\Delta}{\phi}_{\rm photon}~=~2 {\pi}.
   \label{eq:reunion}
\end{equation}
Assuming this and substituting equations (\ref{eq:phi_part} - \ref{eq:pt})
into equation (\ref{eq:reunion}), we can choose a value of $R$ and solve
equation (\ref{eq:reunion}) numerically for $R_0$. We give some values
obtained in this way in the table \ref{tab:pt}.

\begin{table}
   \center
   \begin{tabular}{|c|c|c|}
      \hline
      $R/M$&$R_0/M$&${\Delta}{\tau}/M$\\
      \hline
      $6$&$3.46$&$18.6$\\
      $10$&$3.55$&$31.9$\\
      $20$&$3.58$&$57.8$\\
      $100$&$3.56$&$228$\\
      $1000$&$3.54$&$2038$\\
      \hline
   \end{tabular}
   \caption{Values for a null geodesic leaving a particle in a
   circular geodesic orbit in Schwarzschild spacetime at radius $R$ and
   returning to the particle. $R_0$ is the minimal radial coordinate value of
   the null geodesic (minimum coordinated distance from the centre of the
   black hole) and ${\Delta}{\tau}$ is the proper time the particle has to wait for the
   null geodesic to return.}
   \label{tab:pt}
\end{table}

There are a number of noteworthy features of table \ref{tab:pt}. First, at 
at approximately $R_0=3.2M$, the angular deflection of the null geodesic is
$2{\pi}$, and this therefore represents the value of $R_0$ corresponding to a
particle orbit at $R{\rightarrow}{\infty}$.  Next, we note that the values of ${\Delta}{\tau}$ take values
between ${\sim}3R$ for the closest orbits and ${\sim}2R$ for the most distance orbits,
asymptoting to ${\Delta}{\tau}=2R$ in the limit $R{\rightarrow}{\infty}$. In other words, for orbits at large 
distances the time is dominated by the time for the photon to reach the black 
hole and return.

The ${\Delta}{\tau}$'s quoted in table \ref{tab:pt} represent approximate upper bounds on
the extent of the normal neighbourhood along the past worldline of the
particle since they demarcate two intersections of the particle's geodesic
with a null geodesic. Thus, for a particle at a fixed radius $R$, we need not
worry about the convergence of the expansion of the quasi-local part of the
self-force beyond ${\sim}2R$ to the past. 

Let us now consider two such expansions.  First, we consider the expansion for
a \textit{static} particle at radius $R$ coupled to a minimally-coupled
massless scalar field by scalar charge $q$. We take the expression for
$V(x,x')$ given by Anderson and Hu\cite{AndHu} and take partial derivatives
with respect to the Schwarzschild coordinates. Next, we convert these
coordinates into proper time using the coordinate parameterization
\begin{equation}
   {\Delta}{\phi}=0,~ ~ ~{\Delta}{\theta}=0,~ ~ ~{\Delta}r=0,~ ~ ~{\Delta}t=\sqrt{1-\frac{2M}{r}}({\tau}-{\tau}'),
\end{equation}
which is appropriate for a static particle.  We also, without loss of
generality, set ${\theta}={\pi}/2$. We can then integrate the expansion of ${\nabla}_{\alpha} V(x,x')$
with respect to proper time ${\tau}$ to obtain
\begin{eqnarray}
   \fl f^r_{\rm QL} = \frac{9}{2240} \frac{q^2 M^2}{R^{15}}\left(4R-11M\right)
   \left(R-2M\right)^5{\Delta}{\tau}^5 \nonumber \\
   \lo+\frac{1}{3360}\frac{q^2M^2}{R^{20}}\left(20R^3-195 M
   R^2+598 M^2R-585 M^3\right)\left(R-2M\right)^6{\Delta}{\tau}^7\label{eq:QLSF_stat}\\
   \lo+O({\Delta}{\tau}^8),\nonumber
\end{eqnarray}
which is the only non-vanishing component of the quasi-local part of the
self-force to the order of this expansion.

Denote the fifth and seventh order terms in equation (\ref{eq:QLSF_stat}) as
\begin{eqnarray} \fl f^r_{\rm QL}[5]{\equiv}\frac{9}{2240} \frac{q^2
   M^2}{R^{15}}\left(4R-11M\right) \left(R-2M\right)^5{\Delta}{\tau}^5,\\
   \fl f^r_{\rm QL}[7]{\equiv}\frac{1}{3360}\frac{q^2M^2}{R^{20}}\left(20R^3-195 M
   R^2+598 M^2R-585 M^3\right)\left(R-2M\right)^6{\Delta}{\tau}^7.
\end{eqnarray}
Then the fractional truncation error which is induced by only taking terms in
the series expansion of the quasi-local part of the self-force to order ${\Delta}{\tau}^7$
can be estimated as
\begin{equation}
   {\varepsilon} {\equiv} \frac{f^r_{\rm QL}[7]}{f^r_{\rm QL}[5]+f^r_{\rm QL}[7]}.
\end{equation}
This gives an estimate of the upper limit on the \textit{local} truncation
error (the error in truncating the next term in the series) rather than the
more desirable \textit{global} truncation error (the error in truncating all
remaining terms in the series). Nonetheless, it is a standard measure of
truncation error for this kind of analysis where neither the exact solution
nor a form for the general term in the series is known, and is in any case the
best estimate of truncation error we have for this series.

\begin{figure} 
   \begin{center} 
      \epsfxsize=10cm
      \epsfbox{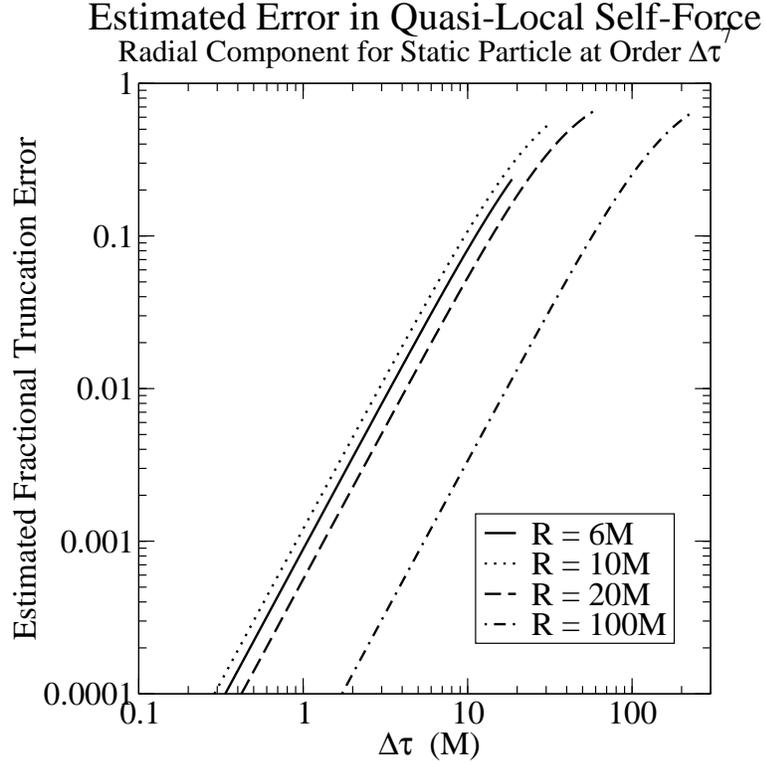} 
   \end{center}
   \caption{
   \label{fig:statr}
   Estimated fractional error ${\varepsilon}$ as a function of ${\Delta}{\tau}$ in the radial
   component of the self-force on a static charge coupled to a massless
   minimally-coupled scalar field in a Schwarzschild spacetime. We show a 
   particle at four different distances from the black hole. Upper bounds for
   the range of ${\Delta}{\tau}$ are those calculated in table \ref{tab:pt}.
   } 
\end{figure}

As noted previously, the truncation error can be expected to grow with ${\Delta}{\tau}$.
We have calculated the error as a function of ${\Delta}{\tau}$ for particles located at
$r=6M$, $10M$, $20M$ and $100M$. The results are presented in figure
\ref{fig:statr}.
We see that, despite our concern that the quasi-local expansion might fail to
converge at the boundary of the normal neighbourhood, it does, in fact, appear
to converge everywhere within the normal neighbourhood. Furthermore, it
converges quite well for particles at all radii, with estimated fractional
error less than 0.1, back to almost ${\Delta}{\tau}=10M$. For $R=6M$, which is the closest
particle that we consider and also the case for which the self-force should be
most important in calculating templates for LISA, the fractional error is less
than $0.1$ more than half way out to the boundary of the normal neighbourhood.

We can apply exactly the same method to a particle in a circular geodesic
orbit at radius $R$ around Schwarzschild. In this case, the coordinates are
related to proper time by
\begin{equation}
   \fl
   {\Delta}{\theta}=0,~ ~ ~{\Delta}r=0,~ ~ ~{\Delta}{\phi}=\frac{1}{R}\sqrt{\frac{M}{R-3M}}({\tau}-{\tau}'),~ ~ ~
   {\Delta}t=\sqrt{\frac{R}{R-3M}}({\tau}-{\tau}').
\end{equation}
Again, we have set ${\theta}={\pi}/2$. In this case, there are three non-vanishing
components of the quasi-local part of the self-force
\begin{eqnarray}
   \fl f^t_{\rm QL} = \frac{q^2M^3}{R^{12}} \sqrt{\frac{R}{R-3\,M}}
      \frac{1}{\left( R-3\,M \right)^3}\nonumber\\
   \lo{\times}\Biggl[\frac{3}{2240}R^3
      \left(R-2\,M\right)\left(R-3\,M\right) 
      \left(5\,R-19\,M\right){\Delta}{\tau}^4\nonumber\\
   +\frac {1}{13440}\, 
      \left(56\,R^4-728\,R^3M
      +3461\,R^2M^2-6990\,M^3R+5073\,M^4\right){\Delta}{\tau}^6 \nonumber\\
   +O({\Delta}{\tau}^7)\Biggr] \\
   \fl f^r_{\rm QL} = \frac{q^2M^2}{R^{14}}
      \frac{\left(R-2\,M\right)}{\left(R-3\,M \right)^{3}}\nonumber\\
   \lo{\times}\Biggl[\frac{9}{11200}\, R^3 \left(R-3\,M \right)
      (20\,R^3-189\,MR^2+558\,M^2R-523\,M^3) {\Delta}{\tau}^5\nonumber\\
   +\frac{1}{47040}\,(\begin{array}[t]{l}
         \phantom{\Biggl(}
         280\,R^5-4690\,R^4M+30780\,R^3M^2 \\
         -97302\,M^3R^2+147777\,RM^4-86481\,M^5){\Delta}{\tau}^7
         \end{array}\nonumber\\
   +O({\Delta}{\tau}^8)\Bigg]\\
   \fl f^{\phi}_{\rm QL} = -\frac{q^2M^2}{R^{13}}\sqrt{\frac{M}{R-3\,M}} 
      \frac{\left(R-4\,M\right)\left(R-2\,M\right)}{\left( r-3\,M \right)^3}
      \nonumber\\
   \lo{\times}\Biggl[\frac {9}{2240}\, R^3 \left(R-3\,M\right)
      \left(3\,R-7\,M\right){\Delta}{\tau}^4\nonumber\\
   +\frac{1}{13440}\,\left(70\,R^3-685\,MR^2+1968\,M^2R-1731\,M^3\right)
      {\Delta}{\tau}^6\nonumber\\
   +O({\Delta}{\tau}^7)\Biggr]
\end{eqnarray}

\begin{figure} 
   \begin{center} 
      \epsfxsize=10cm
      \epsfbox{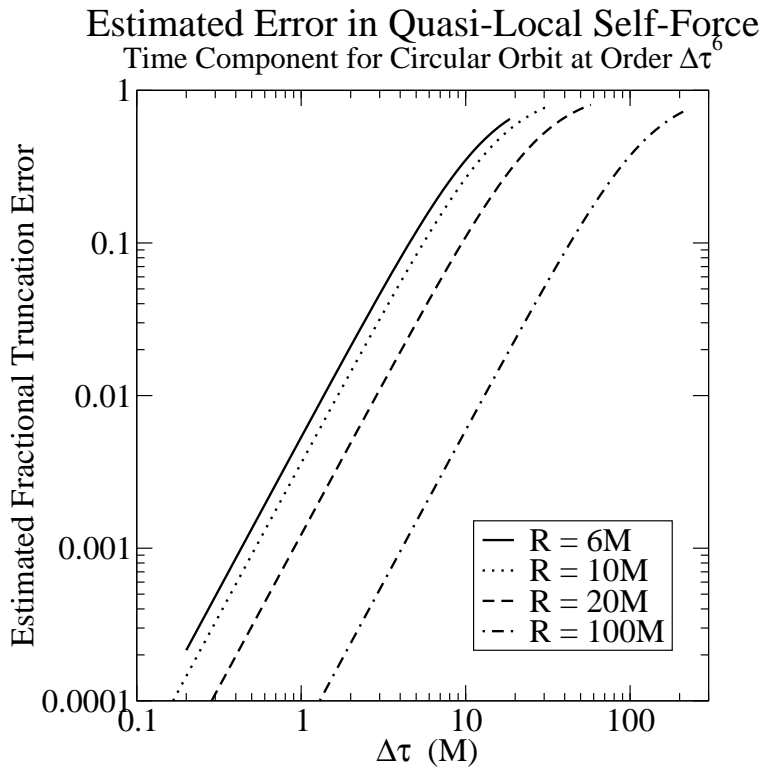} 
   \end{center}
   \caption{
   \label{fig:circt}
   Estimated fractional error as a function of ${\Delta}{\tau}$ in the time component of
   the self-force on a charge coupled to a massless minimally-coupled scalar
   field orbiting on a circular geodesic in a Schwarzschild spacetime. We show
   a particle at four different distances from the black hole. Upper bounds
   for the range at each distance are again approximately those from table
   \ref{tab:pt}.
   } 
\end{figure}

\begin{figure} 
   \begin{center} 
      \epsfxsize=10cm
      \epsfbox{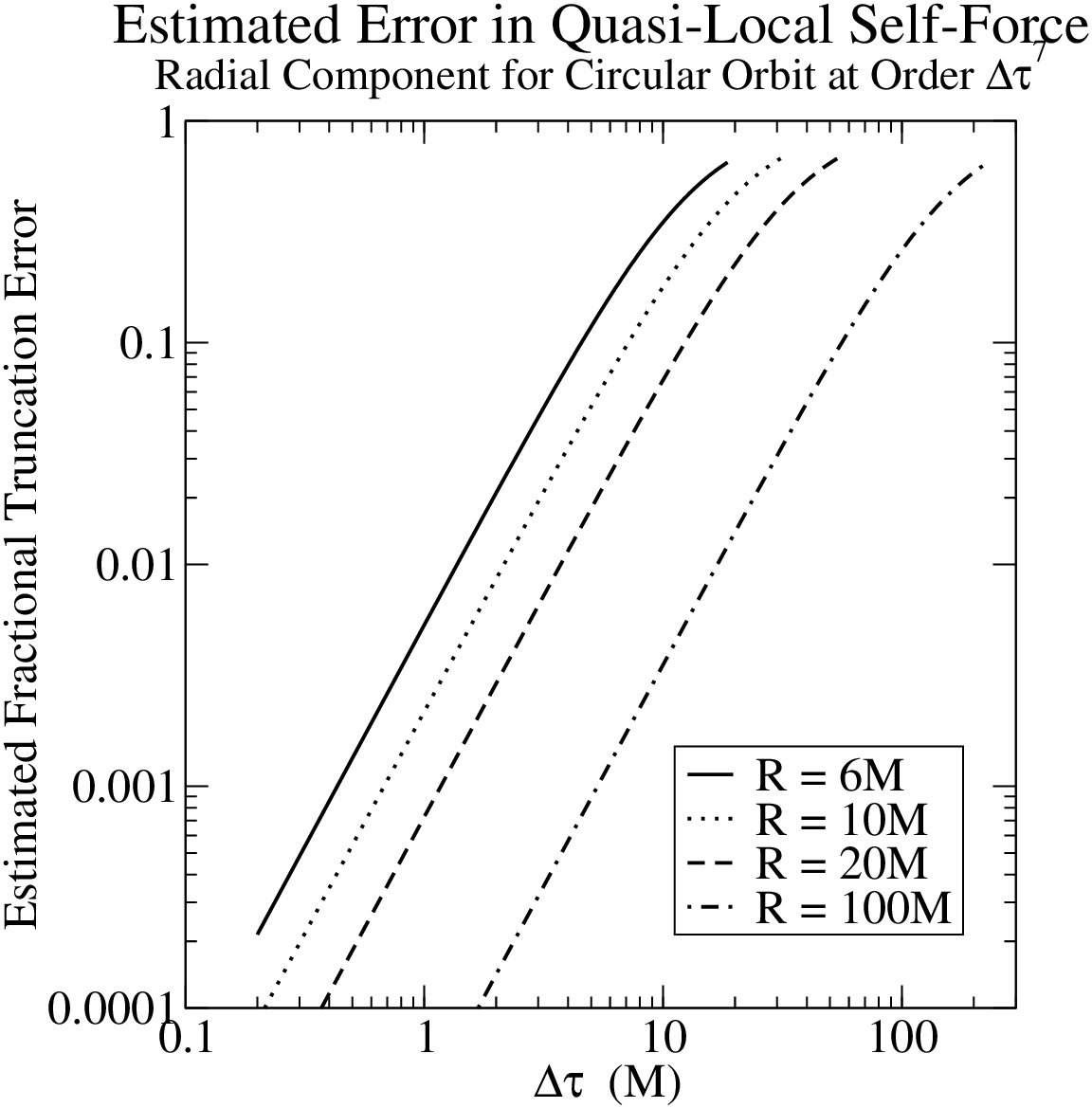} 
   \end{center}
   \caption{
   \label{fig:circr}
   Estimated fractional error as a function of ${\Delta}{\tau}$ in the radial component of
   the self-force on a charge coupled to a massless minimally-coupled scalar
   field orbiting on a circular geodesic in a Schwarzschild spacetime. We show
   a particle at four different distances from the black hole.
   } 
\end{figure}

\begin{figure} 
   \begin{center} 
      \epsfxsize=10cm
      \epsfbox{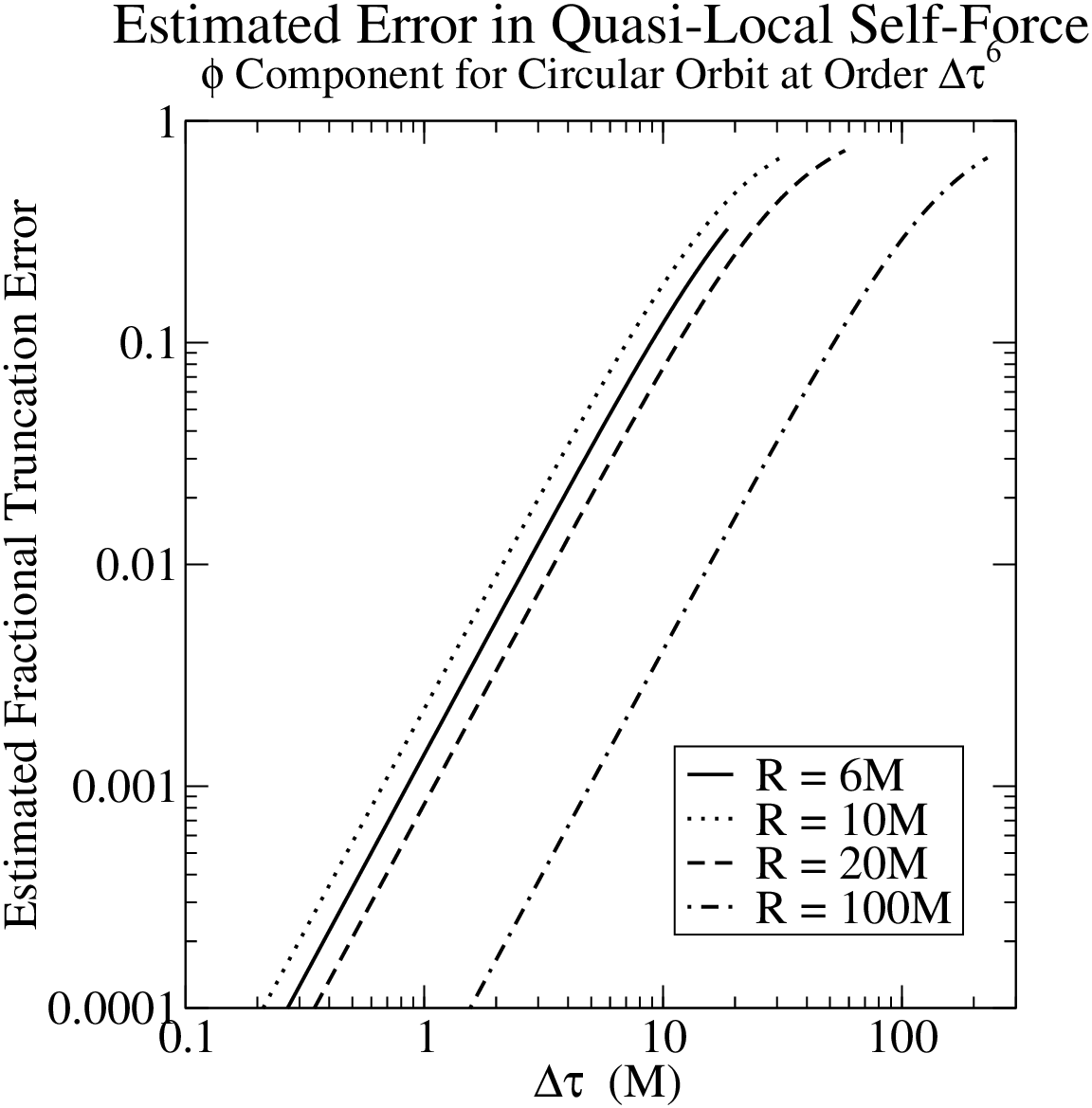} 
   \end{center}
   \caption{
   \label{fig:circphi}
   Estimated fractional error as a function of ${\Delta}{\tau}$ in the ${\phi}$ component of
   the self-force on a charge coupled to a massless minimally-coupled scalar
   field orbiting on a circular geodesic in a Schwarzschild spacetime. We show
   a particle at four different distances from the black hole.
   \ref{tab:pt}.
   } 
\end{figure}

We compute the estimated fractional error for each of these components in the
same manner as we did for the static particle.  The results are presented in 
figures
\ref{fig:circt}, \ref{fig:circr} and \ref{fig:circphi}.
Again, we see convergence everywhere within the normal neighbourhood for all
components of the quasi-local part of the self-force, and good convergence up
to fairly high values of ${\Delta}{\tau}$.

\section{The contribution to the self-force from the distant
past}\label{sec:Wiseman}

In the previous section, we examined the contribution to the self-force from
the portion of the worldline within the normal neighbourhood of the field
point  at $z({\tau})$, i.e. from the first integral in equation (\ref{eq:PWSF}).  In this
section, we will examine the contribution from the earlier part of the
trajectory.  In particular, we will examine the feasibility of computing the
second integral in equation (\ref{eq:PWSF}) using a mode sum expansion for the
Green's function.


\begin{figure} 
   \begin{center} 
      \epsfxsize=12cm
      \epsfbox{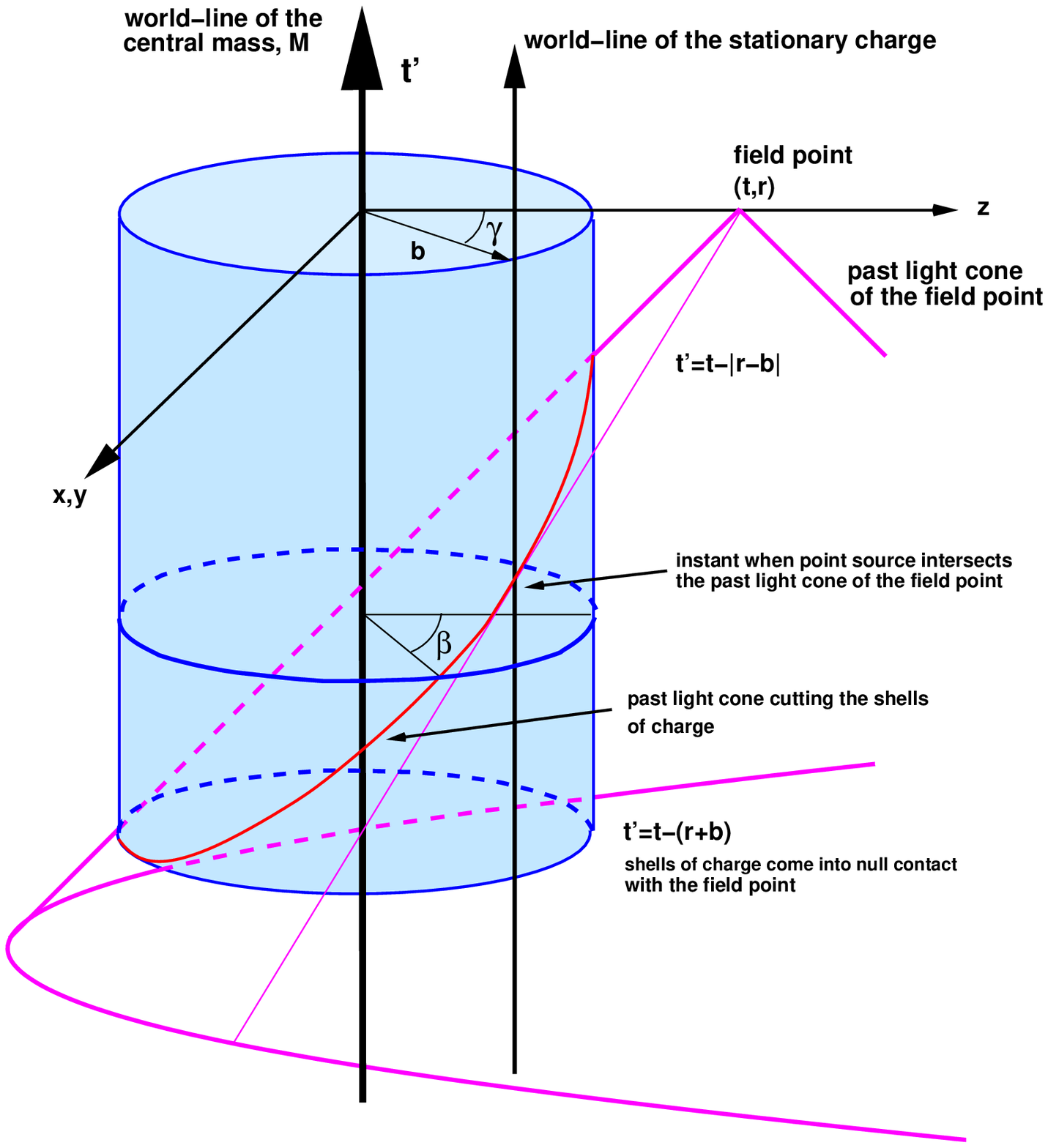} 
   \end{center}
   \caption{
   \label{fig:causal}
   The point charge only has an instant of causal contact with
   the field point $(t,r)$ when the past light cone intersects
   the worldline of the charge; however when using multipoles,
   the point charge is smeared over a sphere and therefore has an
   extended period of interaction.  In equation (\ref{eq:fsGradient}) we have placed
   the field point on the worldline of the source, 
   i.e. ${\gamma}=0$ and $r=b$.
   In  figure \ref{fig:normGradient} we show how the
   gradient of the field ``builds up" for a partial sum of multipoles as we
   integrate over $t^{\prime}$. 
   In computing the numbers in table \ref{tab:converge}, we have again  placed
   the field point on the worldline of the source and we have integrated 
    from ${\beta}= {\pi}$ to ${\beta}= {\pi}/2$, i.e. 
    $\cos {\beta}{\in}[-1,0]$. 
   } 
\end{figure}

\subsection{ The O[M] Green's function}

In studying the forces on a freely falling electric charge in a
curved spacetime, DeWitt and DeWitt \cite{DD64} developed some
very clever techniques for finding the Green's function for a
spherically symmetric spacetime with mass $M$ at the centre.
They give an approximate expression -- accurate to leading order
in the central mass -- for the  Green's function for an electric
charge in this Schwarzschild-like spacetime.  This method was later
extended by Wiseman \cite{wisemanAtCapra99} and  Pfenning and
Poisson \cite{PP00} to the case of a scalar charge. The Green's
function is found to be
\begin{eqnarray}
\fl
G(x,x^{\prime}) = 
{ {\delta}[t-t^{\prime}-|{\bf x} - {\bf x^{\prime}} |] \over |{\bf x} - {\bf x^{\prime}} | } 
\nonumber \\
\fl \;\;\;
+ M {{\partial}\over {\partial}t^{\prime}} 
\left\{
2 { {\delta}[t-t^{\prime}-|{\bf x} - {\bf x^{\prime}} |] \over |{\bf x} - {\bf x^{\prime}} | } 
\ln \left( {r+r^{\prime}+ (t-t^{\prime})  \over r+r^{\prime}- (t-t^{\prime})  } \right)
- 4 { {\Theta}[ (t-t^{\prime})- (r+ r^{\prime})  \,]  \over   (t-t^{\prime})^2 - |{\bf x} - {\bf x^{\prime}} |^2 }
\right\}  \nonumber \\
+O[M^2] \;,
\label{eq:OMGFanalytic}
\end{eqnarray}
where $(t,{\bf x})$ is the field-point  and $(t^{\prime},{\bf x^{\prime}})$ is the source
point.  The first term is clearly identifiable as the retarded Green's
function in Minkowski (flat) spacetime. The logarithmic term is a correction
to the light cones\footnote{This term reflects the fact that the light cones are
bent in curved spacetime.
It was not included in the DeWitt-DeWitt \cite{DD64} calculation.}.  The
final term is the ``tail" of the Green's function. Notice that this term only
contributes prior to $t^{\prime}= t-(r+r^{\prime})$, the light reflection time.  This is
depicted in figure \ref{fig:merged}.  This gives a strong indication that the
self-force is dependent on the portion of the worldline that is outside the
normal neighbourhood.

Multiplying by spherical harmonics and integrating over the solid
angle, we can obtain the angular mode decomposition of this
Green's function
\begin{eqnarray}
G(x,x^{\prime}) = && 
\sum_{l=0}^{\infty} {(2l+1) \over 2 r r^{\prime}} 
\left\{ 
\left[
1 + 2M  {{\partial}\over {\partial}t^{\prime}} 
\ln \left( {r+r^{\prime}+ (t - t^{\prime}) \over r+r^{\prime}- (t - t^{\prime}) } \right)
\right] \right. \nonumber \\
&& {\times}
P_l({\xi}) \; {\Theta}[ t-t^{\prime}- |r- r^{\prime}|  \,] \;
{\Theta}[  r+ r^{\prime}-  ( t-t^{\prime})  \,]
\nonumber \\
&&
\left.
+ 4 M {{\partial}\over {\partial}t^{\prime}}
Q_l({\xi}) \; {\Theta}[ t+t^{\prime}- (r+ r^{\prime}) ]
\right\} \; P_l[\cos({\gamma})] \nonumber \\
&&+O[M^2] \;\; , \;
\label{eq:OMGFmodes}
\end{eqnarray}
where  $P_l$ and $Q_l$ represent Legendre functions and 
\begin{eqnarray}
&& {\xi}= {r^2 +{r^{\prime}}^2 - ( t-t^{\prime})^2 \over  2  r r^{\prime}} = \cos {\beta}\\
&&\cos {\gamma}= \cos {\theta}\, cos {\theta}^{\prime}+\sin {\theta}\, \sin {\theta}^{\prime}\, \cos({\phi}- {\phi}^{\prime})
\end{eqnarray}
and $(t,r,{\theta},{\phi})$ and
$(t^{\prime},r^{\prime},{\theta}^{\prime},{\phi}^{\prime})$ are the
spherical coordinates of the field-point and source-point respectively.
The angle ${\beta}$ is shown in figure \ref{fig:causal}.  In
equation (\ref{eq:OMGFmodes}), it is understood that the first partial
derivative not only operates on the natural log, but also on
$P_l({\xi})$ and the ${\Theta}$-functions.

In numerical implementations, we would only be able to 
include a finite number of multipoles.  In this case,
we can avoid doing a numerical summation by computing the 
partial sum analytically, i.e.
\begin{eqnarray}
\fl
G(N;t,r,t^{\prime},r^{\prime},\cos {\gamma}) =  \sum_{l=0}^{N} 
\mbox{terms in equation (\ref{eq:OMGFmodes})} \nonumber
\\
\fl
= {1\over 2 r r^{\prime}}
\left\{ 
\left[
1 + 2M 
{{\partial}\over {\partial}t^{\prime}}
\ln \left( {r+r^{\prime}+ (t - t^{\prime}) \over r+r^{\prime}- (t - t^{\prime}) } \right)
\right] \right. \nonumber \\
\fl
{\times}\left[ (N+1)
{ P_{N+1}({\xi}) P_N(\cos {\gamma}) - P_N({\xi}) P_{N+1} (\cos {\gamma})  \over
{\xi}- \cos {\gamma}}
\right]
{\Theta}[ t-t^{\prime}- |r- r^{\prime}|  \,] \;
{\Theta}[  r+ r^{\prime}-  ( t-t^{\prime})  \,] \nonumber \\
\fl
\left.
+4 M {{\partial}\over {\partial}t^{\prime}}
\left[
{ 1 + (N+1) 
\left( P_{N+1}(\cos {\gamma}) Q_N({\xi}) - P_{N+1}(\cos {\gamma}) Q_{N+1} ({\xi}) \right)
\over {\xi}-\cos {\gamma}}
{\Theta}[ t-t^{\prime}- (r+ r^{\prime})  \,] \;
\right] \right\} \nonumber  \\
+O[M^2] \;\;.
\label{eq:OMGFsummed}
\end{eqnarray}
As above, the first partial derivative acts on the natural log as well
as the Legendre functions and the ${\Theta}$-functions.

This formula can now be used in the second integral in
equation (\ref{eq:PWSF}). As in Section \ref{sec:Anderson}, we
focus
our attention on a static source at radius $R$ and $\cos{\gamma}=
1$ (see figure \ref{fig:causal}).  
Notice that the final term -- the tail term --  is a total
derivative. When this is substituted into equation (\ref{eq:PWSF}), we
can integrate by parts and see immediately that the tail
contribution to the force vanishes.  This is just a
re-confirmation of the well known result that there is no
self-force on a static scalar charge in Schwarzschild spacetime
(c.f. Wiseman \cite{Wiseman00}).  
By the same
argument, the natural log term also gives no contribution to the
force integral.

Substituting the remaining term -- the flat-space portion of equation
(\ref{eq:OMGFsummed}) -- into equation (\ref{eq:PWSF}), a straightforward
calculation reveals that the first $N$ multipoles of the source give a radial
component of the force
\begin{eqnarray}
f_r({\Delta}{\tau}, N) =
- {q^2 \over R^2 u^t }
{ (N+1) \over 4 \sqrt{2} }
{\int}_{-1}^{\cos {\beta}_{\rm max}}
{ P_{N+1}({\xi}) -P_{N}({\xi}) \over (1 -{\xi})^{3/2} }  d{\xi}\;,
\label{eq:fsGradient}
\end{eqnarray}
where $u^t = 1 -M/b +O[M^2]$ is the leading order contribution to the
time-component of the four-velocity of the static charge. The upper limit is
defined as
\begin{eqnarray}
\cos {\beta}_{\rm max} {\equiv}  1 - {1\over 2} \left( { {\Delta}{\tau}\over  u^t \, b} \right)^2.
\end{eqnarray}
As can be seen from figure \ref{fig:causal}, when the field point is on the
worldline of the charge, $\cos{\beta}$ can be used to parameterise the portion of
the worldline from the field point ($\cos{\beta}=1$) backwards to the time when the 
past null cone of the field point first intersects the cylinder around the 
central mass on which the worldline lies ($\cos{\beta}=-1$). As we will discuss
below, this is the only portion of the worldline which can contribute to the
self-force in this case. In our case, we do not integrate to the field point,
but rather to some time ${\Delta}{\tau}$ to the past of it as per figure \ref{fig:merged},
thus terminating our the integral at $\cos{\beta}_{\rm max}$. This is indicated on
the left hand side of equation (\ref{eq:fsGradient}) by the argument ${\Delta}{\tau}$.

The only portion of the Green's function that contributed to equation
(\ref{eq:fsGradient}) was the flat-space term.  Since the static point
particle would never intersect the past light cone of a field point located on
its own worldline, we can ask: why doesn't $f_r({\Delta}{\tau}, N) $ vanish identically?
The answer lies in the fact that we are using only a finite number of modes to
describe the field gradient.

The decomposition of the field into angular modes can also be thought of
as a mode decomposition of the source. In the case of the static
point particle, we are, in effect, replacing the point charge
with a sum of spherical shells of charge where each
shell has radius $R$ and an angular distribution of charge
smeared on it. The point particle is only recovered when a large
(infinite) number of shells of charge are summed. As is evident 
from figure \ref{fig:causal}, although the point charge doesn't
come into null contact with the source point, the spherical shells
of charge do intersect the past light cone of the field point.
If we terminate the integral at some ${\Delta}{\tau}>0$, but
include an infinite number of terms in the summation (i.e.
$N{\rightarrow}{\infty}$), we will recover the point-particle nature
of the source, and therefore there would be no contribution to the
force.  However, in any practical calculation, we will need to
terminate at some finite value of $N$, and thus we will be left
with an unwanted, and inescapable, contribution to the force.
The key question is can we include enough terms in the summation
to squeeze this unwanted contribution to an acceptably low level?

\begin{figure} 
   \begin{center} 
      \epsfxsize=8cm
      \epsfig{file= 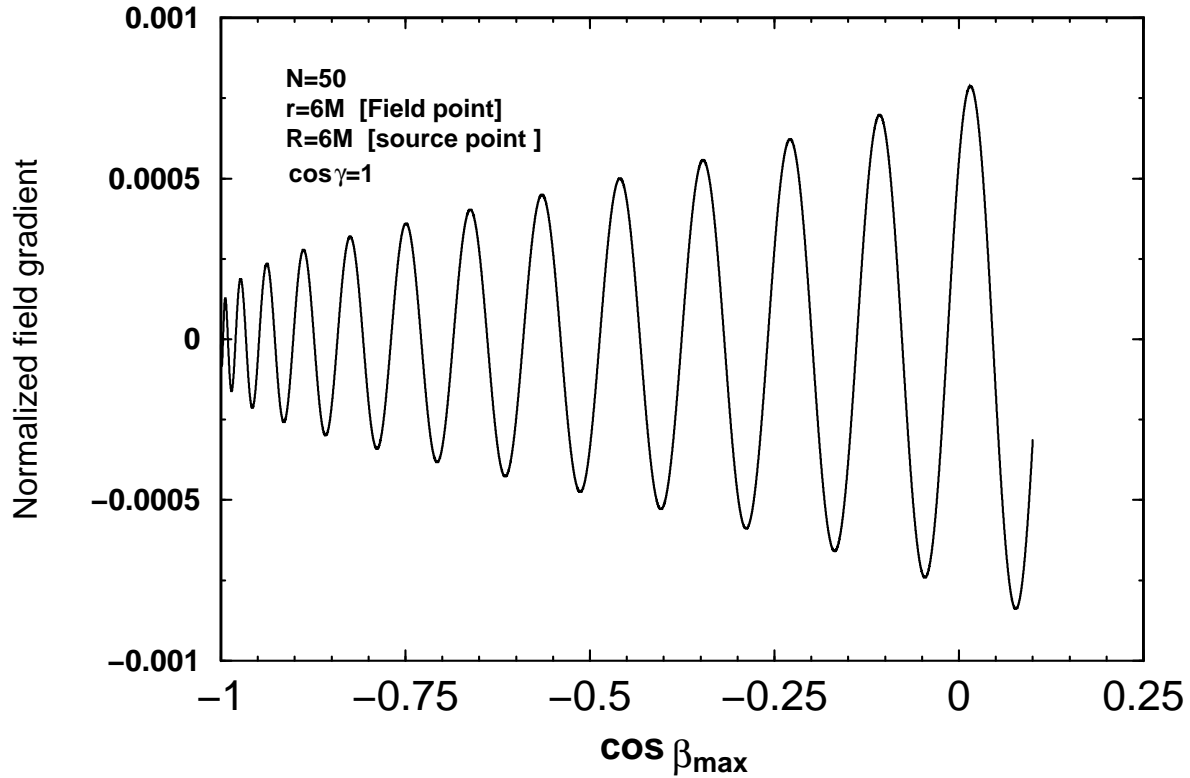, angle=-90, width=\linewidth} 
   \end{center}
   \caption{
   \label{fig:normGradient}
   This shows the accumulation of the field gradient as we
integrate over ${{\tau}^{\prime}}$. The normalization is $(M^2 u^t/q^2)
f_r({\Delta}{\tau}, N)$. Here $N$ is chosen to be $50$. If we had chosen
a larger $N$,  the frequency of oscillation in the plot would increase, but  
the  envelope of the oscillation would become smaller.  The numbers in the table
\ref{tab:converge} are a measure of how fast the envelope shrinks
as the number of modes is increased. For example the entry for 50 modes 
is just the value of the peak near $\cos {\beta}_{\rm max}=0$. 
   } 
\end{figure}

Figure \ref{fig:normGradient} shows how the field gradient accumulates as we
integrate over ${{\tau}^{\prime}}$ for a fixed number of modes. The oscillations are
artefacts of the finite number of modes summed, and it is the envelope of the
function which reflects the level of accuracy that can be achieved. If we are
to accurately evaluate the value for the self-force, we must squeeze the
envelope of this function below the value of the self-force. For example, on
dimensional grounds, one might expect there to be a radial component of the
self-force for a static scalar charge in Schwarzschild spacetime of the form
\begin{eqnarray}
f_r = {\lambda}{q^2 M \over b^3 } \; , 
\label{eq:hypotheticalforce}
\end{eqnarray}
where ${\lambda}$ is an unknown coefficient we would be trying to dig out from under
the unwanted contribution to the force from our finite mode sum (in the
electrostatic case, there is a finite contribution to the self-force of
exactly the form shown in equation (\ref{eq:hypotheticalforce}) with ${\lambda}=1$).
The amplitude of the envelope is therefore an indication of the bound we place
on the contribution to the self-force. In the present case, we know the
solution vanishes identically, and evaluating equation (\ref{eq:QLSF_stat})
from section \ref{sec:Anderson} at $R=6M$, it can be seen that the first
integral in equation (\ref{eq:PWSF}) gives only a tiny contribution  to the
self-force of a static scalar charge, even when ${\Delta}{\tau}$ is extended out to near
the edge of the normal neighbourhood.  Therefore, in this case, the amplitude
of the envelope is nothing but the approximate accuracy to which we have
calculated the self-force.

If we choose ${\Delta}{\tau}$ such that $\cos {\beta}_{\rm max} =
0$, which is about half-way out to our upper bound on the edge of
the normal neighbourhood, then the we see from table
\ref{tab:converge} that using 200 multipoles we can constrain the
${\lambda}$ to be less than $0.1$.

\begin{table}
   \center
   \begin{tabular}{|c|c|c|}
      \hline
      Modes & $(M^2u^t/q^2)f_r({\Delta}{\tau}, N)$ &  bound on ${\lambda}$      \\
      \hline
      $10$      &   $0.0017 $   &   $0.36 $   \\
      $50$      &   $0.00078$   &   $0.17 $  \\
      $100$     &   $0.00056$   &   $0.12 $  \\
      $200$     &   $0.00041$   &   $0.088$  \\
      $300$     &   $0.00036$   &   $0.078$  \\
      \hline
   \end{tabular}
   \caption{Table showing the convergence of the mode sum and the
    accuracy to which we could use it to compute the gradient
    appearing in the integrand of equation (\ref{eq:PWSF}). The
    first column is the number of modes included in the sum. 
    The integral is terminated at ${\Delta}{\tau}$ chosen such that
    ${\beta}_{\rm max} = 0$. The charge is located at $R=6M$.
    This value in the second column is the peak nearest 
    $\cos{\beta}_{\rm max} =0$ on plots 
    similar to figure \ref{fig:normGradient}. 
    The right column shows the accuracy with which you could
    constrain the coefficient of a self-force of the form 
    ${\lambda}q^2 M/b^3$ with the number of modes used.
}
\label{tab:converge}
\end{table}

\section{Conclusions} \label{sec:Conc}

Our goal in this paper was to assess the feasibility of the Poisson-Wiseman
matched expansion scheme for calculating the self-force. This scheme involves
splitting the non-local term in the self-force (which is an integral over the
particles worldline $z({\tau})$) at some proper time ${\tau}-{\Delta}{\tau}$. For the $-{\infty}<{\tau}'<{\tau}-{\Delta}{\tau}$
(distant) part of the integral, the integrand is expanded as a mode sum.  For
the ${\tau}-{\Delta}{\tau}<{\tau}'<{\tau}$ (quasi-local) part, the integrand is expanded in a covariant
Taylor series, which requires $z({\tau}-{\Delta}{\tau})$ be in the normal neighbourhood of
$z({\tau})$.  For the resulting self-force to be accurate, the split must be done
in such a way that each expansion reach sufficient accuracy with a finite and
feasible number of terms. Until now, it has not been clear that such a split
even existed.

For this preliminary investigation, we have studied the simplest scenarios:
minimally-coupled massless scalar field, Schwarzschild geometry, static
particles and circular geodesic motion.  Our results are somewhat surprising.
The quasi-local expansion appears to converge well - when the particle is at
$6M$ and expanding to order ${\Delta}{\tau}^7$, we achieve an estimated truncation error
of the order of a few percent for ${\Delta}{\tau}{\sim}6M$. This would na\"{i}vely have seemed
to us enough of a buffer between the mode-sum integral and the singularity to
allow rapid convergence of the mode sum as well.

However, we have found the convergence of the mode sum to be poor. This is
because the mode sum smears the charge of the particle over spheres of finite
radius which extend outside of the normal neighbourhood. Thus, even the
flat-space term in the Green's function, which cannot contribute to the
self-force for a scalar particle, seems to contribute at every order. For our
simple case of a static particle at $6M$ and ${\Delta}{\tau}{\sim}6M$, going from 10 modes to
100 increased our accuracy by only a factor of 3.

Clearly, we need a way to speed convergence of mode sum. In the case
presented, one could have achieved infinite accuracy at every order by
regularizing the modes, which would have removed the flat-space portion of the
Green's function. It might, therefore, be possible to speed the convergence of
the mode sum by regularizing the modes\cite{RegModes}. This would be ironic,
since this method is supposed to provide an alternative to mode
regularization. Nonetheless, it might be that combining regularization with a
two expansion approach will provide much better convergence than either alone.
Alternatively, we note in figure \ref{fig:normGradient} that the mode-sum
integrand oscillates about its true value. This leads us to speculate that
averaging over these oscillations would lead to much quicker convergence of
the mode-sum integral. 

While we believe our convergence analysis is general, and will apply beyond
the simple scenarios we have explored, we would be remiss in not pointing out
that there are additional issues for higher spin fields. The foremost of these
is the issue of gauge - in order to meaningfully combine the quasi-local and
more distant self-force contributions, they will need to be expressed in the
same gauge. For instance, the quasi-local contribution for the gravitational
field\cite{AFO} has only be derived in the Lorentz gauge, where no expression
for the modes is available.  We note, however, that this is a generic
difficulty for gravitational self-force calculations because the formal
expressions for the self-force are themselves derived in the Lorentz
gauge\cite{MST97,QW97}, and we are heartened by current progress in
understanding gauge transformations in this context\cite{BO01}. 

In any case, we feel that these preliminary results are promising enough to
warrant further investigation. A good figure of merit for calculational
schemes like this is accuracy per floating point operation. As is, this
approach would seem to lag approaches like mode-sum regularization when
measured on this scale. Nonetheless, it might, for instance, provide
confirmation for results from other approaches. Further, as mentioned above,
it is still possible that this calculational scheme, with refinements, could
rival or exceed in computational efficiency other known schemes, especially
since the quasi-local expansion, once calculated, can be applied with little
further effort to any spacetime or particle motion. In the mean time, this
approach remains, in our opinion, in the category of ``promising but
possessing some technical challenges''.
\bigskip

\ack 
This work was supported primarily by NSF grants PHY-0140326 and PHY-0200852,
with additional support  by the Center for Gravitational Wave Physics, which
is funded by the National Science Foundation under Cooperative Agreement PHY
0114375, the Center for Gravitational Wave Astronomy, which is funded by
the National Aeronautic and Space Administration through the NASA University
Research Center Program, and the Yukawa Institute for Theoretical Physics. 
We would like to thank the Capra Gang, and especially Eric Poisson, for useful
conversations. We would also like to thank the organizers of the Capra
Meeting series for providing excellent forums for the exchange of ideas and
information.


\begin{thebibliography}{8}

\bibitem{LISA} See http://lisa.jpl.nasa.gov/

\bibitem{Ryan96} F. D. Ryan, Phys. Rev. D \textbf{53}, 3064 (1996).

\bibitem{Dirac38} P. A. M. Dirac, Proc. Roy. Soc. \textbf{A167}, 148
(1938).

\bibitem{Poisson99} E. Poisson, {\it An introduction to the
Lorentz-Dirac equation}, gr-qc/9912045.

\bibitem{DB60} B. S. DeWitt and R. W. Brehme, Ann. Phys. (N.Y.) \textbf{9},
   220 (1960).

\bibitem{Hobbs68} J. M. Hobbs, Ann. Phys. \textbf{47}, 141 (1968).

\bibitem{MST97} Y. Mino, M. Sasaki and T. Tanaka, Phys.\ Rev.\ D \textbf{55}, 
   3457 (1997). 

\bibitem{QW97} T. C. Quinn and R. M. Wald, Phys.\ Rev.\ D \textbf{56} 3381,
   (1997).

\bibitem{Quinn00} T.C. Quinn, Phys.Rev. D \textbf{62}, 064029 (2000).

\bibitem{Poisson03} E. Poisson, {\it The motion of point particles in
  curved spacetime}, gr-qc/0306052, to appear in Living Reviews in Relativity. 

\bibitem{Rorlich63} F. Rohrlich, Ann. Phys. (N.Y.) \textbf{22} 169, (1963).

\bibitem{Boulware80} D.G. Boulware, Ann. Phys. (N.Y.) \textbf{124} 169, (1980).

\bibitem{Matsas94} G. E. A. Matsas, Gen. Rel. Grav. \textbf{26} 1165, (1994).

\bibitem{Hadamard} J. Hadamard, \textit{Lectures on Cauchy's Problem in
   Linear Partial Differential Equations} (Yale University Press, New Haven,
   1923).

\bibitem{Gunther} Articles in Zeitsch.  Analys. Anwend., 16, (1997). 

\bibitem{B&D} N. D. Birrell and P.C.W. Davies, \textit{Quantum fields in
   curved space} (Cambridge University Press, Cambridge, England, 1982).

\bibitem{PW97} E. Poisson and A.G. Wiseman, suggestion at the First 
   Capra Ranch meeting on Radiation Reaction, Capra Ranch, 1998.

\bibitem{C76} S. M. Christensen, Phys.\ Rev.\ D \textbf{14}, 2490 (1976).

\bibitem{C78} S. M. Christensen, Phys.\ Rev.\ D \textbf{17}, 946 (1978).

\bibitem{Roberts} M. D. Roberts, Class. Quantum Grav. \textbf{6}, 419-423,
   (1989).

\bibitem{AndHu} P. R. Anderson and B. L. Hu, Phys. Rev. D. {\bf 69}, 064039
   (2004).

\bibitem{AFO} W. G. Anderson, \'{E}. \'{E}. Flanagan and A. C. Ottewill, {\it
   Quasi-local contribution to the gravitational self-force}, Phys, Rev. D
   \textbf{71}, 024036 (2005).

\bibitem{DD64} C. M. DeWitt and B. S. DeWitt, Physics (N.Y.) \textbf{1}, 3
   (1964). 

\bibitem{wisemanAtCapra99} A.G. Wiseman, presentation at the
Second Capra Ranch meeting on Radiation Reaction, Dublin, Ireland, 1999,
unpublished.

\bibitem{PP00} M. J. Pfenning and E. Poisson, \textit{Scalar, electromagnetic,
   and gravitational self-forces in weakly curved spacetimes}, gr-qc/0012057.

\bibitem{Wiseman00} A. G. Wiseman, Phys. Rev. D \textbf{61}, 084014 (2000).

\bibitem{RegModes} Place holder for special issue articles on mode
   regularization.

\bibitem{BO01} L. Barack and A. Ori, Phys. Rev. D \textbf{64}, 124003 (2001).
%
%
%
%
%
%
%
%
%
%
%
%
%
%
%
%
%
%
%
%
%
%
%
%
%
%
%
%
%
%
%
%
%
%
%
%
%
%
%
%
%
%
%
%
%
%
%
%
%
%
%
\end{thebibliography}
\end{document}